\documentclass[aps,prb,showpacs,floatfix,superscriptaddress,twocolumn]{revtex4-1}
\usepackage{graphicx}    
\usepackage{hyperref}    
\usepackage{bm}          	
\usepackage[sumlimits,intlimits]{amsmath}
\usepackage{amsfonts,amssymb}
\usepackage[section]{placeins}

\begin{document}

\title{Excitons in anisotropic 2D semiconducting crystals}

\author{A. S. Rodin}
\affiliation{Boston University, 590 Commonwealth Ave., Boston MA 02215}
\author{A. Carvalho}
\affiliation{Graphene Research Centre and Department of Physics, National University of Singapore, 117542, Singapore}
\author{A. H. Castro Neto}
\affiliation{Boston University, 590 Commonwealth Ave., Boston MA 02215}
\affiliation{Graphene Research Centre and Department of Physics, National University of Singapore, 117542, Singapore}

\date{\today}
\begin{abstract}
The excitonic behavior of anisotropic two-dimensional crystals is investigated using numerical methods. We employ a screened potential arising due to the system polarizability to solve the central-potential problem using the Numerov approach. The dependence of the exciton energies on the interaction strength and mass anisotropy is demonstrated. We use our results to obtain the exciton binding energy in phosphorene as a function of the substrate dielectric constant.
\end{abstract}


\maketitle
\section{Introduction}

The field of two-dimensional (2D) crystals has been undergoing a rapid development since the famous isolation of graphene~\cite{Novoselov2004eff}. Over the years, new materials have been added to the catalogue of 2D systems, such as boron nitride, silicene, and a variety of transition metal dichalcogenides. The most recent addition to this growing family is black phosphorus. This material is composed of individual phosphorene layers, held together by the van der Waals force. The weak interlayer force makes it possible to separate the bulk into few-layer structures~\cite{Li2014bpf,Liu2014pau,Qiao2014flb,Xia2014rbp,Koenig2014eff,Castellanos_Gomez2014iac,Buscema2014fab}. In addition, a recent study has been published demonstrating a technique of obtaining monolayer phosphorene~\cite{Lu2014paf}.

Despite being a fairly recent addition to the 2D library, phosphorene exhibits a number of features that set it apart from other members and make it attractive for the physics community. First, with the exception of graphene, phosphorene is the only 2D system composed of a single type of atoms. Unlike graphene, however, phosphorene has a gap which is sensitive to the mechanical deformation of the lattice and the number of layers~\cite{Rodin2014sig,Qiao2014flb,Castellanos_Gomez2014iac,Tran2014tbg,Fei2014set,Peng2014sed}. Another trait that distinguishes phosphorene is its high anisotropy, leading to a highly asymmetric band structure. The existence of the tunable gap makes phosphorene an interesting material in the context of excitons. However, the complex electronic structure makes the study rather difficult. Work has been done on determining the binding energies of excitons in black phosphorus using first-principles calculations~\cite{Tran2014tbg} and variational methods~\cite{Castellanos_Gomez2014iac}. However, there have been no systematic studies of the excitonic behavior and its dependence on the variable system parameters such as the band structure and the strength of interaction. In this paper, we address this problem using numerical methods. We begin by deriving a general expression for the potential inside a polarizable 2D system in the presence of a bulk dielectric. Following this, we adopt several simplifications to reduce the computation time. Finally, we obtain the dependence of the excitonic energy levels on the system anisotropy and the interaction strength. The results obtained here are applicable for both direct- and indirect-gap systems as it is the curvature of the bands which is important for determining the binding energies. The consequence of the indirect gap is a longer excitonic lifetime due to the momentum mismatch between the conduction and valence band extrema. Therefore, our analysis applies broadly to a variety of gapped 2D systems and make it attractive for basic science and applications.

\section{Dielectric Screening}
It is known that the Coulomb interaction in thin dielectric sheets has a nontrivial form due to screening~\cite{Keldysh1979cii,Cudazzo2011dsi,Berkelbach2013ton,Castellanos_Gomez2014iac}. Strictly speaking, the Keldysh interaction~\cite{Keldysh1979cii} applies to thin layers of finite thickness. Since the concept of thickness is ill-defined for single layers, one should be careful when using this particular result. Earlier work~\cite{Cudazzo2011dsi} has obtained the modified Coulomb interaction for a 2D sheet in vacuum. Incidentally, it has the same functional form as the Keldysh interaction, but the system parameters have different origins. Here, we extend the earlier result by adding a bulk dielectric positioned at distance $h$ below the 2D sheet to function as a substrate. Keeping $h$ finite allows one to study suspended samples.

Our system consists of a dielectric slab with susceptibility $\chi$ located at $z < 0$ and a two-dimensional layer situated at $z = h$. We position a charge $q$ at $\mathbf \rho_0 = (0,0,h)$ and calculate the potential it creates within the layer. From the Poisson's equation, we have
\begin{equation}
\frac{-\nabla^2\Phi}{4\pi} =q\delta^3\left(\mathbf \rho - \rho_0\right)+\delta(z-h)\sigma_L(r)+\delta(z)\sigma_B(r)\,,
\label{eqn:Poisson}
\end{equation}
where $r$ is the planar coordinate, $\sigma_L(r)$ is the charge density in the 2D layer, $\sigma_B(r)$ is the bound surface charge on the bulk dielectric, and $\Phi$ is the total potential. It is convenient to take the Fourier transform of this expression:
\begin{equation}
\frac{\left(p^2+k^2\right)\hat\Phi}{4\pi} =\frac{qe^{i h k}}{(2\pi)^{3/2}}+\mathcal{F}_z\left[\delta(z-h)\tilde\sigma_L+\delta(z)\tilde\sigma_B\right]\,.
\label{eqn:Fourier_Eqn}
\end{equation}
We use a hat to denote the 3D transform and a tilde for the 2D planar transform. $\mathbf p$ labels the in-plane momentum and $\mathbf k$ is the momentun in $z$-direction. Using the fact that $\sigma_B(r) = \chi E_z(r,z=0)$, we write
\begin{equation}
\sigma_B(r) = \frac{-\chi}{1+2\pi\chi}\left[q\delta^2(r)+\sigma_L(r)\right]\ast\frac{h}{\left(r^2+h^2\right)^\frac{3}{2}}\,,
\label{eqn:sigma_B}
\end{equation}
where asterisk represents the convolution operation. Note that $E_z$ includes the contribution from the point charge, the induced charge in the thin sheet, and the surface charge of the bulk dielectric. Planar Fourier transform of $\sigma_B(r)$ is obtained from the convolution theorem:
\begin{equation}
\tilde\sigma_B =- \frac{2\pi \chi}{1+2\pi\chi}\left[\frac{q}{2\pi}+\tilde\sigma_L\right]e^{-h p}\,.
\label{eqn:sigma_B_F}
\end{equation}

Next, we determine $\sigma_L$. The charge on the 2D sheet  arises as a response to the in-plane field. The polarization is given by $\mathbf P = -\tensor{\zeta}\nabla_p\Phi(r,z=h)$ and $\sigma_L = -\nabla\cdot\mathbf P$, yielding
\begin{equation}
\sigma_L = \left.\zeta_{xx} \Phi_{xx} + \zeta_{yy}\Phi_{yy}+2\zeta_{xy}\Phi_{xy}\right|_{z=h}\,,
\label{eqn:sigma_L}
\end{equation}
where the subscripts on $\Phi$ label the partial derivatives. We also set $\zeta_{xy} = \zeta_{yx}$. This allows us to write
\begin{align}
\tilde\sigma_L &= -R(p)\underbrace{\int\frac{\hat\Phi}{\sqrt{2\pi}}e^{-ihk'}dk'}_{\tilde \Phi_{2D}(p)}\,,
\label{eqn:sigma_L_F}
\\
R(p)& = \zeta_{xx}p_x^2 + \zeta_{yy}p_y^2+2\zeta_{xy}p_xp_y\,.
\end{align}
Plugging Eqs.~\eqref{eqn:sigma_B_F} and \eqref{eqn:sigma_L_F} into Eq.~\eqref{eqn:Fourier_Eqn}, one obtains
\begin{align}
\tilde\Phi_{2D}(p)&=\frac{qS(p)}{1+2\pi R(p)S(p)}\,,\quad S = \frac{1-\frac{\epsilon-1}{\epsilon+1}e^{-2hp}}{p}\,,
\label{eqn:Phi_2D_F}
\end{align}
where we have used $\epsilon = 1+4\pi\chi$.To make the expression in Eq.~\eqref{eqn:Phi_2D_F} more amenable to our calculations, we make several simplifications. First, we position the 2D sheet on top of the dielectric, setting $h = 0$. Next, we set $\zeta_{xy} = 0$ and $\zeta_{xx} = \zeta_{yy} = \zeta$. Taking the inverse Fourier transform of the simplified Eq.~\eqref{eqn:Phi_2D_F} gives
\begin{align}
\Phi_{2D}(r)&=\frac{\pi q}{2\kappa r_0}\left[H_0\left(\frac{r}{r_0}\right)-Y_0\left(\frac{r}{r_0}\right)\right]\,.
\label{eqn:Phi_2D}
\end{align}
Here, $H_0(r)$ and $Y_0(r)$ are Struve and Bessel functions, respectively. We have introduced the length scale $r_0 =2\pi\zeta/\kappa$ with $\kappa = (1+\epsilon)/2$. This simplified result reduces to the one obtained in Ref.~\onlinecite{Cudazzo2011dsi} for $\kappa  = 1$.

While it might appear that our $\zeta_{xx} = \zeta_{yy}$ is rather crude, it is possible to replace both by their average provided they don't differ substantially. This will be addressed in the context of phosphorene in a latter section.

\section{Anisotropic Masses}
We now move on to the two-body problem with direction-dependent masses. The center-of-mass Hamiltonian for an anisotropic two-body system with an attractive central potential is given by
\begin{equation}
H = \frac{p_x^2}{2\mu_x}+\frac{p_y^2}{2\mu_y}-V\left(\frac{\mathbf d}{r_0}\right)\,,\quad \mu_{x/y} = \frac{m_{x/y}M_{x/y}}{m_{x/y}+M_{x/y}}\,,
\label{eqn:H_General}
\end{equation}
where $\mathbf d$ is the separation between the particles, $m$ and $M$ are the masses of electrons and holes, and $\mu_{x/y}$ is the direction specific reduced mass. It is more convenient to address this problem by going from anisotropic masses to an anisotropic potential by performing a change of variables
\begin{equation}
\sqrt{\frac{\mu_{x/y}}{2\bar\mu m_e}}d_{x/y} = r_{x/y}\,,\quad \bar\mu = \frac{\mu_x\mu_y}{\mu_x+\mu_y} \frac{1}{m_e}\,.
\label{eqn:Scale_Var}
\end{equation}
This results in
\begin{equation}
H = -\frac{\hbar^2}{4\bar\mu m_e}\nabla^2 -V\left(\frac{r\sqrt{1+\beta\cos2\phi}}{r_0}\right)\,,
\label{eqn:H_Scale_Var}
\end{equation}
with $\beta = (\mu_y-\mu_x)/(\mu_y+\mu_x)$ for $\mu_y>\mu_x$.

A problematic trait of our central potential is its singularity. In addition, the wavefunctions change much more at small $r$, so we need to emphasize them in the solution. Thus, we perform a change of variables $t = \ln (r/r_0)$:
\begin{align}
H &= -\frac{\text{Ha}}{4\bar\mu W^2}\left[e^{-2t}\left(\partial_t^2+\partial_\phi^2\right) +\right.
\nonumber
\\
&+\left.GU\left(e^t\sqrt{1+\beta\cos2\phi}\right)\right] = \frac{\text{Ha}}{GW}\mathcal{H}\,,
\label{eqn:H_Change_Var}
\\
U(y)&=\frac{\pi}{2}\left[H_0\left(y\right)-Y_0\left(y\right)\right]\,,\quad  G =\frac{4}{\kappa^2}\bar\mu W\,,
\label{eqn:Potential}
\end{align}
where $W$ is $2\pi\zeta$ divided by the Bohr radius and Ha is the Hartree energy. The benefit of this transformation turns out to be not only the removal of the singularity, but also of the first derivative, bringing the equation to the appropriate form to be solved by the Numerov method.

\section{Numerical Approach}

Having set up the problem, we proceed to the numeric solution. From Eq.~\eqref{eqn:H_Change_Var}, we are tying to solve the reduced Hamiltonian problem
\begin{equation}
\mathcal H\Psi = \mathcal E\Psi\,,\quad \Psi = \sum _m a_m \Lambda_m\,,
\label{eqn:Schrodinger}
\end{equation}
where $\Lambda_m$ are the basis functions and $a_m$ are their respective coefficients. As expected for a central potential, the general form of a basis function is
\begin{equation}
\Lambda = \sum_{l=0}^\infty  g_l(t)\cos l\phi+ h_l(t)\sin l\phi\,.
\end{equation}
Plugging it into Eq.~\eqref{eqn:Schrodinger}, we write
\begin{widetext}
\begin{equation}
-\sum_{l=0}^\infty\left[e^{-2t}\left(\partial_t^2-l^2\right) +GU\left(e^t\sqrt{1+\beta\cos2\phi}\right)\right]\left[g_l(t)\cos l\phi+ h_l(t)\sin l\phi\right] = \mathcal{E}\sum_{l=0}^\infty\left[g_l(t)\cos l\phi+ h_l(t)\sin l\phi\right]\,.
\label{eqn:Big_Sch}
\end{equation}
\end{widetext}
Because of the $\cos 2\phi$ term in the potential, it only couples sines to sines and cosines to cosines. Moreover, it is clear that not only do sines and cosines couple exclusively among themselves, but also that even and odd angular momentum coefficients to not mix. Thus, because of the harmonic mixing introduced by the anisotropy, eigenstates now fall into one of four classes: $\mathbf c^{e/o}$ and $\mathbf s^{e/o}$, where $\mathbf c$ and $\mathbf s$ label the harmonic function and the superscript designates whether the angular momenta are even or odd. For the isotropic case, $\mathbf c^e$ includes $s$ and $d_{x^2-y^2}$ orbitals, $\mathbf c^o$ contains $p_x$ and $f_{y^3-3xy^2}$, $\mathbf s^e$ has $d_{xy}$, and $\mathbf s^o$ represents $p_y$ and $f_{x^3-3yx^2}$. Once the anisotropy is turned on, the orbitals in each class mix, but for small $\beta$ they retain most of their original shape. Therefore, for the sake of convenience, we will refer to the anisotropic wavefunctions using the isotropic orbital names.

An important consequence of the harmonic mixing has to do with the selection rules for the electric dipole transition between the energy levels. The standard electric dipole perturbation is given by $H_1 \sim \mathbf{\varepsilon}\cdot \mathbf r = r\left(\varepsilon_x \cos\phi+\varepsilon_y\sin\phi\right)$, where $\mathbf\varepsilon$ is the field polarization vector. To determine whether a transition is allowed, the matrix element of $H_1$ for the initial and the final states is computed. From the structure of $H_1$, it is known that the particle can move only between energy levels whose angular momenta differ by one. With the introduction of the four anisotropic classes where each state contains multiple angular harmonics this requirement changes. Now the transitions are allowed between classes which contain harmonics that differ by one. In other words, the transitions between even and odd classes are now allowed and those withing even and odd groups are prohibited. Of course, the rate of the transition depends on the contribution of the ``correct" harmonics to the given states. Nonetheless, for large enough $\beta$'s this mechanism can result in a higher rate than, say, electric quadrupole transitions.

Following the discussion above, we set
\begin{equation}
\Psi^\pm = \sum_l f_l(r)\text{trig}^\pm(l\phi)\,,
\end{equation}
where $l$ runs over the appropriate harmonic numbers and $\text{trig}^\pm$ is cosine or sine, respectively, we multiply Eq.~\eqref{eqn:Big_Sch} by $\text{trig}^\pm(n\phi)$  and integrate to get
\begin{align}
&-e^{-2t}\left(\partial_t^2-n^2\right)f_n(t) -G\sum_l U_{nl}f_l(t) = \mathcal{E}f_n(t)\,,
\\
&U_{nl} =\oint\frac{d\phi}{2^{\delta_{n,0}}\pi}  \,U\left(e^t\sqrt{1+\beta\cos2\phi}\right)\text{trig}^\pm(n\phi)\text{trig}^\pm(l\phi)\,.
\label{eqn:Matrix_Element}
\end{align}

We define a vector function $\mathbf f(t) = [ f_{n_0}(t)\,,f_{n_0+2}(t)\dots f_{n_0-2+2N}(t)\,,f_{n_0+2N}(t)]$ where each entry corresponds to a particular angular harmonic. Naturally, one has to terminate the sum at some harmonic number $n_0 +2N$, resulting in $N+1$ terms in the vector function. Note that $n_0$ can be 0, 1, or 2. $n_0 = 0$ corresponds to even-$n$ cosine-like wavefunctions; $n_0 = 1$ is used for odd cosine- and sine-like functions; $n_0 = 2$ applies to even sine-like functions since $n_0 = 0$ results in vanishing sine terms. We also introduce an angular momentum operator $\mathbf n^2$, where $\mathbf n$ is a diagonal matrix of $n$, and the interaction operator $\mathbf U$ which couples the harmonics in accordance with Eq.~\eqref{eqn:Matrix_Element}. Putting everything together allows us to write
\begin{equation}
\mathbf f''(t) = \mathbf M(t)\mathbf f(t)\,,\quad \mathbf M(t) = \mathbf n^2-e^{2t}\left(G\mathbf U+\mathcal E\right)\,,
\label{eqn:Matrix_Eq}
\end{equation}
The form of Eq.~\eqref{eqn:Matrix_Eq} is precisely what is required for the matrix Numerov method.

The Numerov method entails dividing the range of $t$ into $N_t$ steps of size $\Delta t$ and using the following set of relations to connect $\mathbf f_{j\pm 1}$ (where the subscript labels the $t$-position) to two preceding steps:
\begin{align}
\mathbf P_j & = 1-\Delta x^2\frac{\mathbf M_{j}}{12}\,,
\\
\mathbf f_{j\pm1}&=\mathbf P_{j\pm1}^{-1} \left[(12-10\mathbf P_j)\mathbf f_j-\mathbf P_{j\mp1}\mathbf f_{j\mp1}\right]\,,
\label{eqn:RL_Move}
\end{align}
To use this method, one chooses the initial conditions at $\mathbf f_0$, $\mathbf f_1$, $\mathbf f_{N_t-1}$, and $\mathbf f_{N_t}$. Then, one designates a matching point $t_m$, located between $t_0$ and $t_{N_t}$, and uses Eq.~\eqref{eqn:RL_Move} to approach this matching point from the right and the left. As we are using $N+1$ harmonics in the expansion of the basis functions, we need to have $N+1$ basis functions. These are obtained by setting up different initial conditions at the boundaries so that all basis $\mathbf f$'s are linearly independent at the edges.

One needs to be aware of a numerical problem that may arise. As the integration goes forward, the component of the vector $\mathbf f$ corresponding to the largest harmonic grows exponentially faster than others because of the $n^2$ term in $\mathbf M$. This causes the basis vector functions to lose their linear independence by the time $t_m$ is reached. This can be remedied by using the Riley regularization procedure~\cite{Riley1968vet}. Defining $\mathbf V_n$ as a matrix containing all the vectors $\mathbf f^k_{t_n}$, where $k$ labels the basis vector, we transform all the already-computed $\mathbf V$'s by multiplying them by $\mathbf V_n^{-1}$. One needs to perform this procedure regularly to prevent the exponentially growing vector componen from destroying the linear independence of the basis vectors. In fact, if one is only interested in the energies and not the actual wavefunction, it is possible to apply the regularization procedure only to $\mathbf V_n$ and the previous $\mathbf V$ as only two points are used in the Numerov integration. This can substantially reduce the computation time.

Finally, since all harmonics have to be matched at $t_m$ independently, we have
\begin{align}
\sum_l C_l \mathbf f_{t_m}^{l,R}&= \sum_l D_l \mathbf f_{t_m}^{l,L}\,,
\quad
\sum_l C_l \mathbf{\dot f}_{t_m}^{l,R}= \sum_l D_l \mathbf{\dot f}_{t_m}^{l,L}\,,
\end{align}
where $L$ and $R$ denote left- and right- moving solutions, $C_l$ and $D_l$ are the coefficients of the solutions originating from different initial conditions. This can be rewritten as
\begin{equation}
\text{det}\begin{pmatrix}
\mathbf f^{1,R}_{t_m}&\mathbf f^{2,R}_{t_m}&\dots&\mathbf f^{N,R}_{t_m}&\mathbf f^{1,L}_{t_m}&\mathbf f^{2,L}_{t_m}&\dots&\mathbf f^{N,L}_{t_m}
\\
\mathbf{\dot f}^{1,R}_{t_m}&\mathbf{\dot f}^{2,R}_{t_m}&\dots&\mathbf{\dot f}^{N,R}_{t_m}&\mathbf{\dot f}^{1,L}_{t_m}&\mathbf{\dot f}^{2,L}_{t_m}&\dots&\mathbf{\dot f}^{N,L}_{t_m}
\end{pmatrix} = 0\,.
\label{eqn:Det_Equation}
\end{equation}
By varying the energy parameter $\mathcal E$, one solves the determinant equation using the bisection method.

\section{Results}

\begin{figure}
\includegraphics[width = 3in]{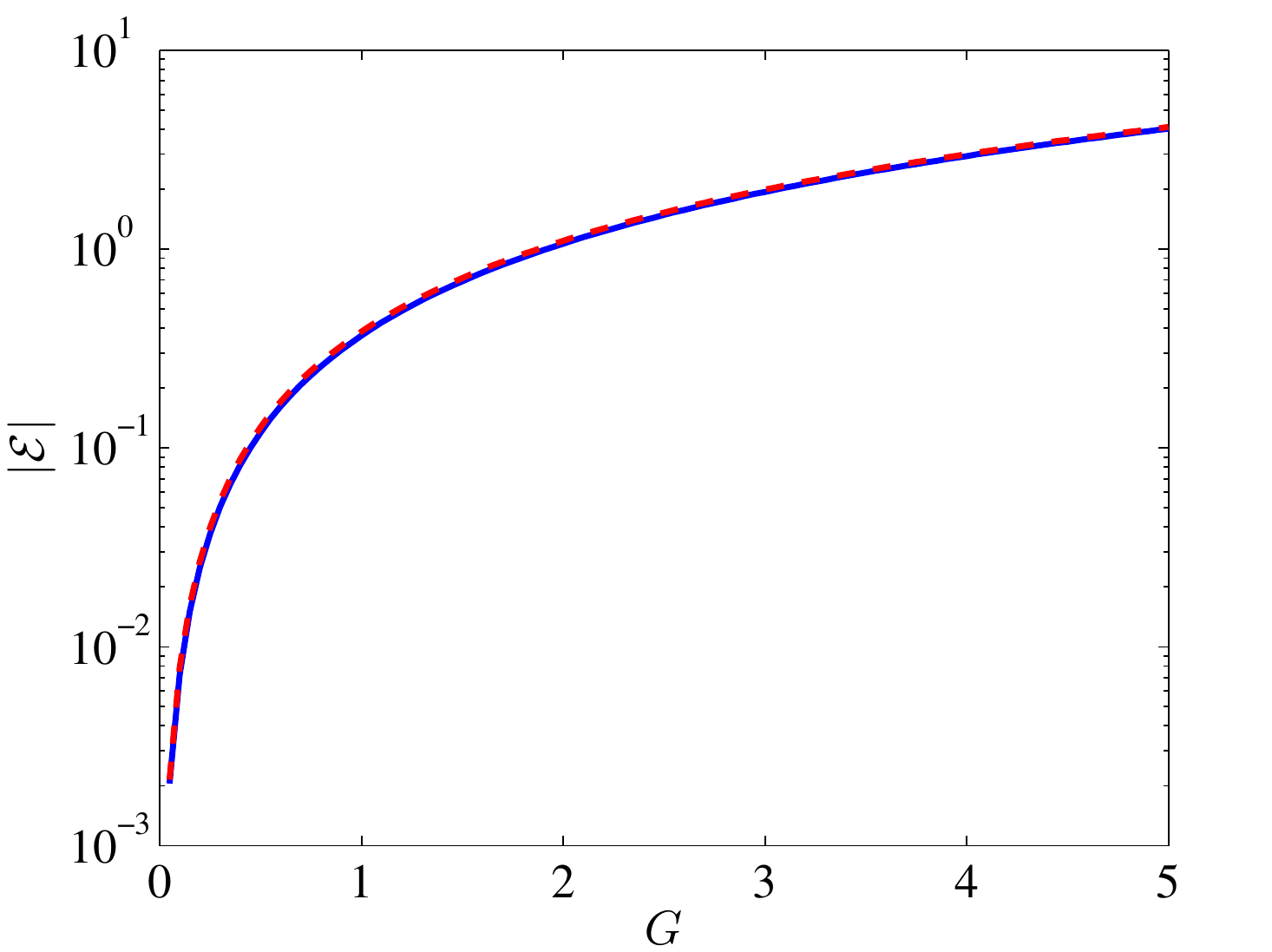}
\caption{(Color online) The ground state energy as a function of $G$ for $\beta = 0$ using the original Eq.~\eqref{eqn:Potential} potential (dashed line) and the simplified version from Eq.~\eqref{eqn:Simp_U} (solid line).}
\label{fig:Interaction_Comparison}
\end{figure}

One downside of the potential in Eq.~\eqref{eqn:Potential} is its complexity as it makes the integral in Eq.~\eqref{eqn:Matrix_Element} rather slow. To speed up the evaluation, we use an approximate form for the potential~\cite{Cudazzo2011dsi}:
\begin{equation}
\bar U(y) = -\left[\ln\left(\frac{y}{y+1}\right)+\left(\gamma-\ln 2\right)e^{-y}\right]\,.
\label{eqn:Simp_U}
\end{equation}
To demonstrate the quality of this simplification, we begin by computing the ground state energies for the isotropic case as a function of $G$ using the original $U$ and the simplified $\bar U$. We use the $\beta =0$ case as it requires no harmonic integration and the $1s$ state can be obtained directly by using a single $l = 0$ harmonic. We present the results in Fig.~\ref{fig:Interaction_Comparison}. As one can see, the agreement is quite good between the two potentials.

Even with the simplified potential, the solution to the problem is still computationally intensive. There are, nevertheless, certain steps that one can take to reduce the time needed to obtain the results. It is clear that the bound-state energies depend on the interaction strength $G$ and the anisotropic parameter $\beta$. However, the coupling matrix $\mathbf U$ depends only on $\beta$. This means that one can fix $\beta$ and calculate $\mathbf U$ once for a particular set of angular harmonics and $t$-grid and then reuse it to obtain energies for different $G$'s. This process can then be repeated for other $\beta$'s and sets of harmonics. As the computation of $\mathbf U$ requires a large number of numerical integrals, doing it only once significantly cuts the computation time.

\begin{figure}
\includegraphics[width = 3in]{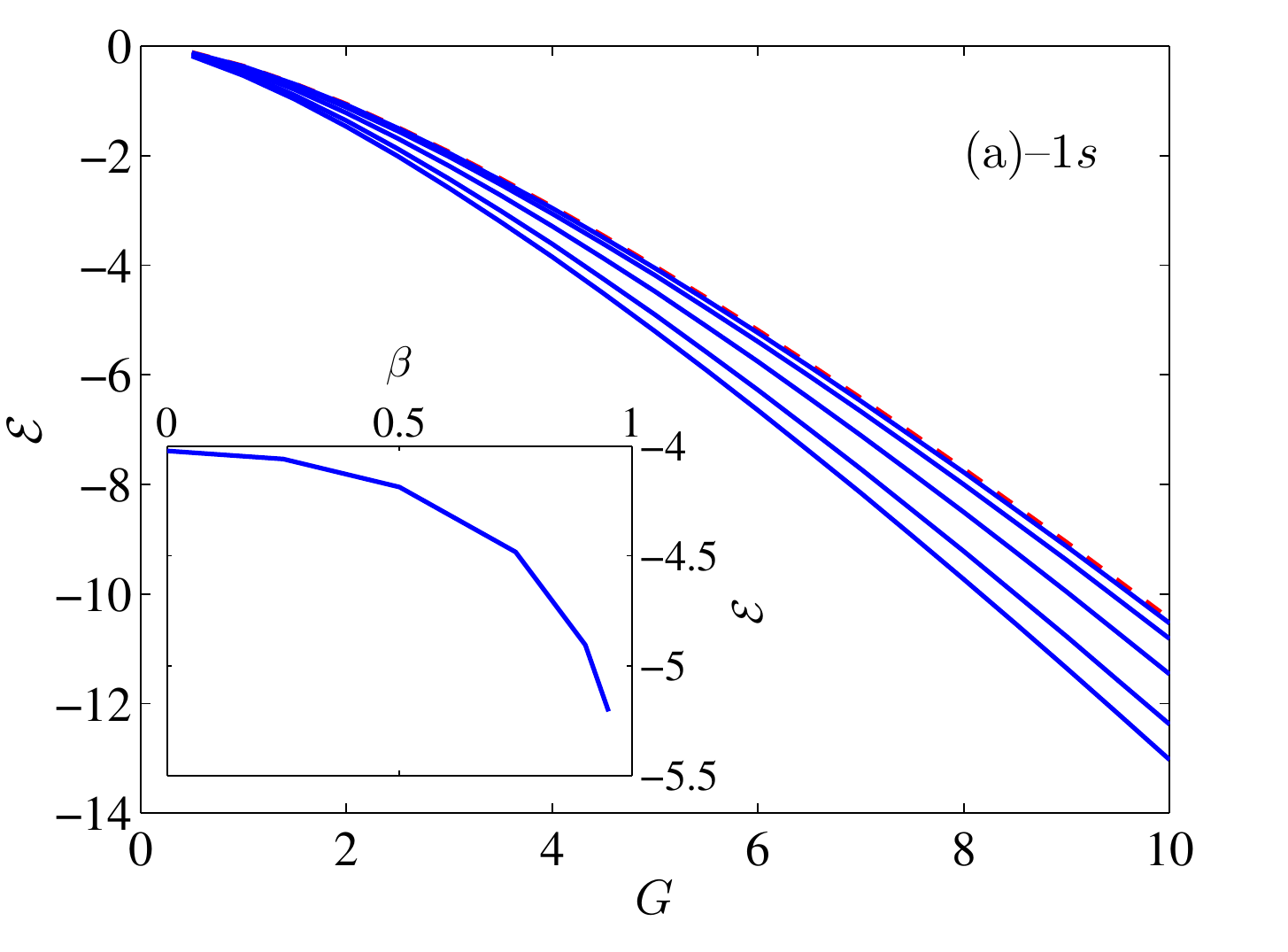}
\includegraphics[width = 3in]{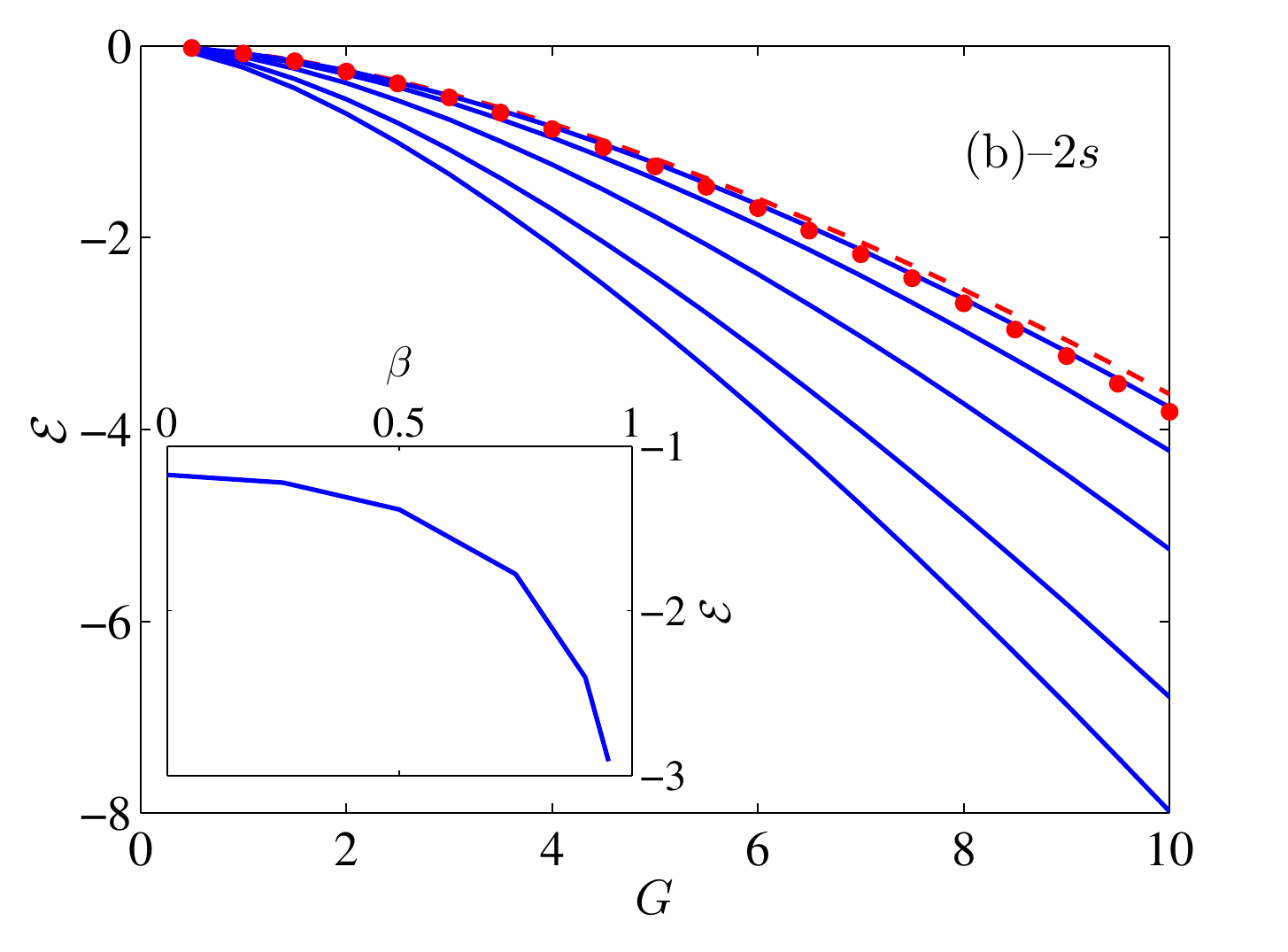}
\\
\includegraphics[width = .8in]{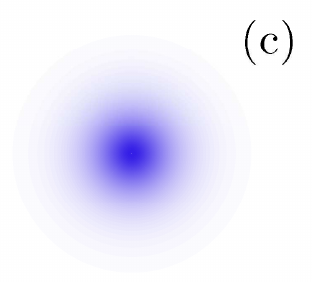}
\includegraphics[width = .8in]{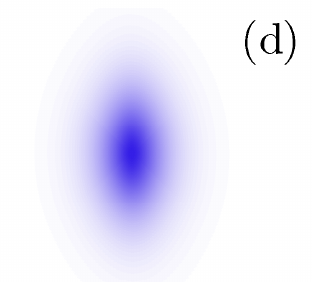}
\includegraphics[width = .8in]{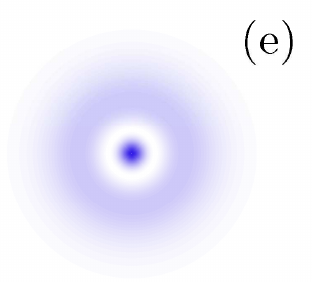}
\includegraphics[width = .8in]{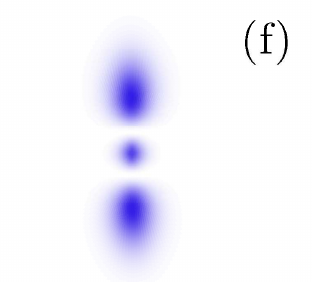}
\caption{$\mathcal E$ versus $G$ for (a) $1s$ and (b) $2s$ orbitals for different $\beta$'s. The dashed lines are $\beta = 0$. Moving from the dashed line down: $\beta = 0.25$, $0.5$, $0.75$, $0.9$, and $0.95$. The circles in (b) are obtained for $\beta = 0$ case using the potential in Eq.~\eqref{eqn:Potential}. The insets show $\mathcal E$ vs. $\beta$ for $G = 5$. (c)--(f) show the probability distribution obtained from the wavefunctions for $\beta = 0$ and $\beta = 0.95$. (c) and (d) correspond to $1s$; (e) and (f) portray $2s$.}
\label{fig:S_Orbitals}
\end{figure}

We are now in the position to perform the necessary calculations. The results for the first two levels of the $s$-like orbital are given in Fig.~\ref{fig:S_Orbitals}. We plot the energies $\mathcal E$ for a range of $\beta$'s to show its dependence on the interaction strength $G$. It is immediately apparent that, while superlinear, $\mathcal E$ changes slower than $G^2$ as it does for the regular Coulomb interaction. One can also see that the $\mathcal E$ changes more rapidly with $G$ for the second energy eigenstate. This means that the relative energy level separation varies with $G$ and cannot be determined from the quantum numbers. Moreover, it is clear that the energy states with higher $\beta$ change with $G$ more than the more isotropic ones. This makes the anisotropic states much more sensitive to the dielectric constant of the bulk dielectric. Comparing the $1s$ and $2s$ states reveals that anisotropy plays a much greater role for the $2s$ orbital. This can be seen by looking at the probability distributions at $\beta = 0.95$. While for $1s$ such a high anisotropy results in a fairly mild deformation from the circularly symmetric case, $2s$ manifests a qualitatively different behavior. The particle cloud outside the orbital node becomes ``folded" into two lobes along the $y$-axis. Analyzing the orbital composition shows that the anisotropic $2s$ case gets its appearance from the combination of the isotropic $2s$ and the $d_{x^2-y^2}$ components. The apparent difference between $1s$ and $2s$, therefore, can be understood in terms of the perturbation theory, regarding the  anisotropic portion of the potential as the perturbation. As $1s$ is the deepest energy state, it is significantly separated from other states with the correct parity in terms of energy. This means that even at larger $\beta$, $1s$ does not pick up a substantial amount of higher-level traits. In contrast, $2s$ is shallower and is located closer to higher-harmonic states, resulting in a greater modification of the wave function.

\begin{figure}
\includegraphics[width = 3in]{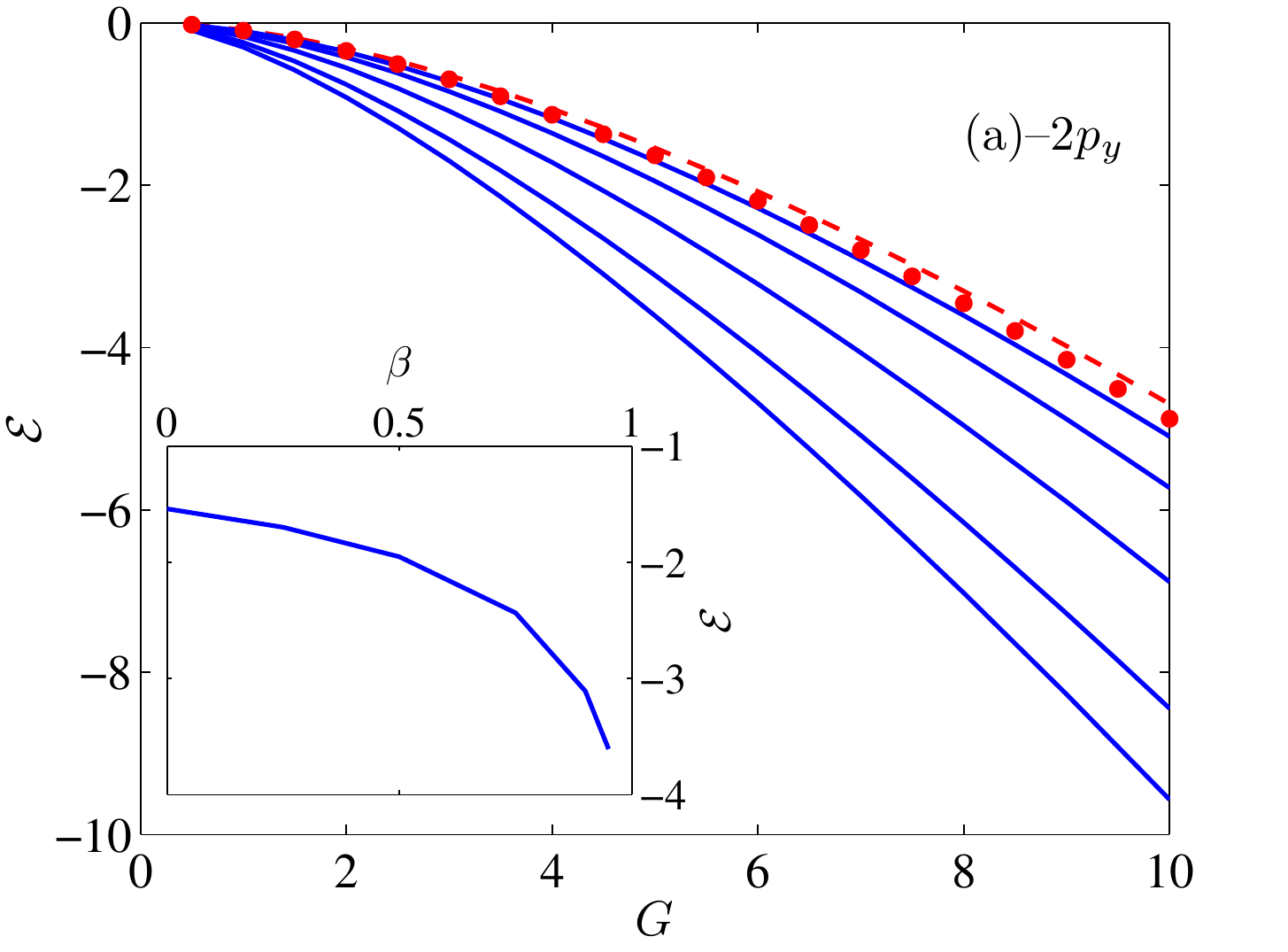}
\includegraphics[width = 3in]{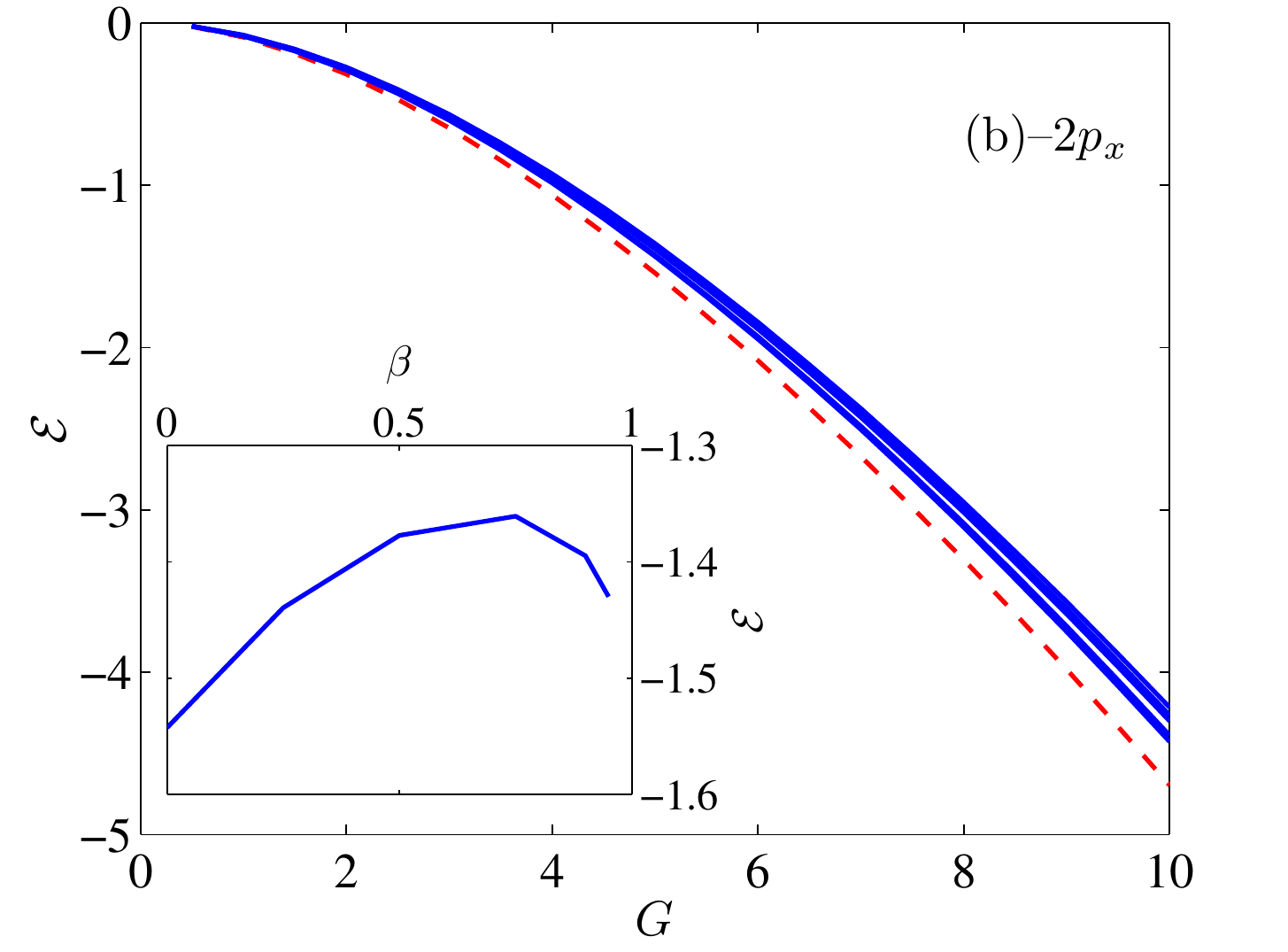}
\\
\includegraphics[width = .8in]{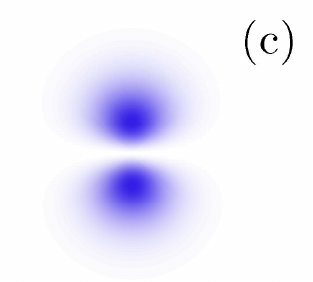}
\includegraphics[width = .8in]{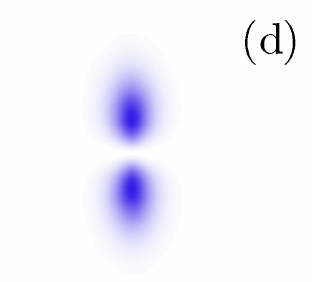}
\includegraphics[width = .8in]{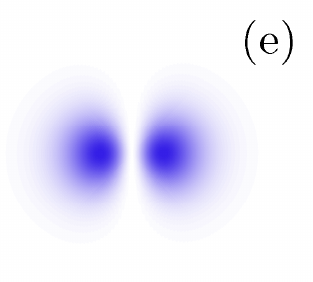}
\includegraphics[width = .8in]{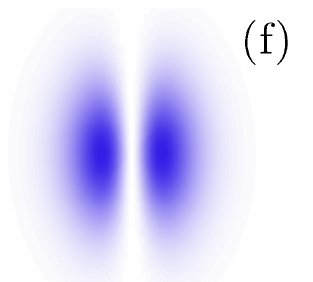}
\caption{$\mathcal E$ versus $G$ for $2p_y$ (a) and $2p_x$ (b) orbitals for the same $\beta$'s as in Fig.~\ref{fig:S_Orbitals}. In (a), lower curves correspont to higher $\beta$. Circles show the $\beta = 0$ for the unsimplified potential. Note the non-monotonicity of $\mathcal E$ for $2p_x$. The insets show the dependence of $\mathcal E$ on $\beta$ for $G = 5$. (c) and (e) show the probability ditributions for $2p_y$ and $2p_x$, respectively, at $\beta = 0$. (d) and (f) show the same for $\beta = 0.95$.}
\label{fig:P_Orbitals}
\end{figure}

Next, we move to the $2p$ orbitals, Fig.~\ref{fig:P_Orbitals}. Here, a stark difference is observed between the $p_x$ and $p_y$ orbitals. $p_y$ demonstrates an expected behavior with $\mathcal E$ becoming more negative at larger $\beta$ and $G$. On the other hand, $p_x$ not only does not depend very strongly on $\beta$, but it also exhibits a non-monotonic variation with the anisotropic parameter. This non-monotonicity has previously been observed in Ref.~\onlinecite{Pfeiffer1993bsf}. To understand this behavior, we need to look at the probability distribution for both orbitals. For $p_y$, the lobes are located along the $y$-axis, which is the direction along which the potential well diverges as $\beta\rightarrow1$. This means that as $\beta$ gets larger, more of the particle cloud experiences the enhanced potential, making $\mathcal E$ more negative. In the case of $p_x$, the lobes are perpendicular to the diverging direction and the wave function actually vanishes along the $y$-axis. Thus, a small anisotropy does not lower the energy of the $2p_x$ orbital, but instead raises it by coupling it to higher energy states. As $\beta$ approaches 1, the potential well gets deeper around the $y$ axis, lowering the energy of the state somewhat. However, since the wave function is still zero along the diverging axis, the energy remains finite.

An important feature of this modified potential is the lifting of the accidental degeneracy. Unlike the standard Coulomb problem, $2s$, $2p_x$, and $2p_y$ all have different energies at finite $\beta$. Of course, $p_x$ and $p_y$ energies coincide at $\beta = 0$, but they are still different from the $s$ orbital. Thus, the energy level picture becomes much richer as the eigenstates of the Hamiltonian separate in the energy space.

\section{Phosphorene}

\begin{figure}[h]
\includegraphics[width = 2.7in]{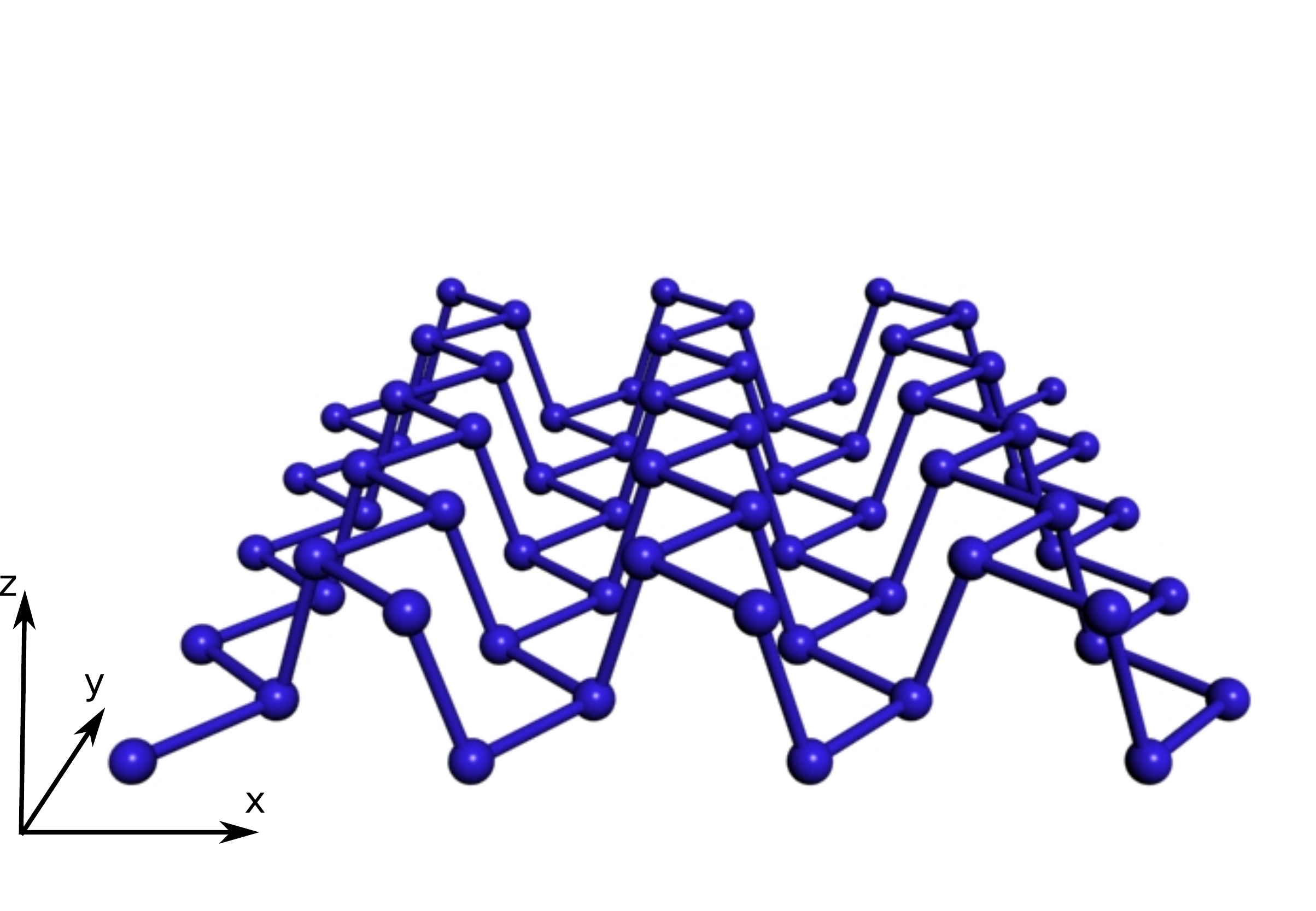}
\\
\includegraphics[width = 3.2in]{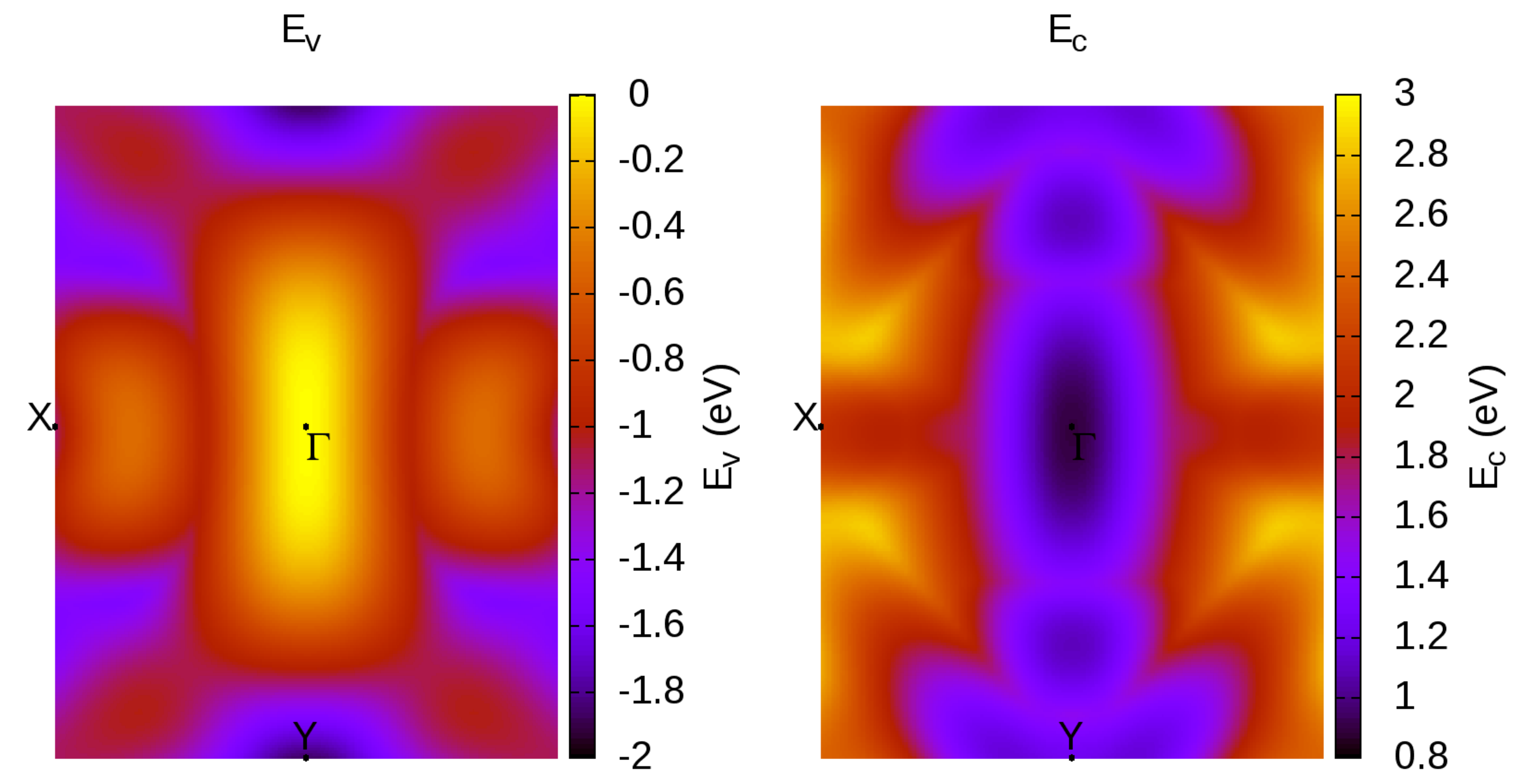}
\caption{Phosphorene lattice and colormaps of the calculated valence and conduction bands. The valence band top is set to zero.}
\label{fig:Map}
\end{figure}

Finally, we address the important case of phosphorene. This phosphorus allotrope is known for its highly anisotropic crystal structure. From the first principles calculations, we obtain the effective electron and hole masses in $x$ and $y$ directions. The band map for the conduction and valence bands is shown in Fig.~\ref{fig:Map}, along with the crystal lattice.

For the electrons, we get $m_x \approx 0.18\pm0.04m_e$ and $m_y\approx 1.23 \pm0.01m_e$. For the holes, $M_x\approx0.13\pm0.04 m_e$ and $M_y$ is a very large number as the band is essentially flat. This yields $\mu_x\approx0.075\pm0.02m_e$ and $\mu_y\approx 1.23m_e$. Using these reduced masses, we obtain $\beta \approx0.89\pm0.02\approx 0.9$ and $\bar\mu\approx 0.7$. To obtain the characteristic length $r_0$, we need the susceptibility of the material.

The 2D susceptibility is obtained using density-functional theory, 
following the method proposed in Ref.~\onlinecite{Berkelbach2013ton},
which is based on the calculation of the dielectric permittivity $\epsilon$
as a function of the interlayer distance ($d$),
\begin{equation}
\epsilon_{x,y}=1+\frac{4\pi\zeta_{xx,yy}}{d}.
\label{eqn:epsilon_xy}
\end{equation}

\begin{figure}[h]
\includegraphics[width =3.1in]{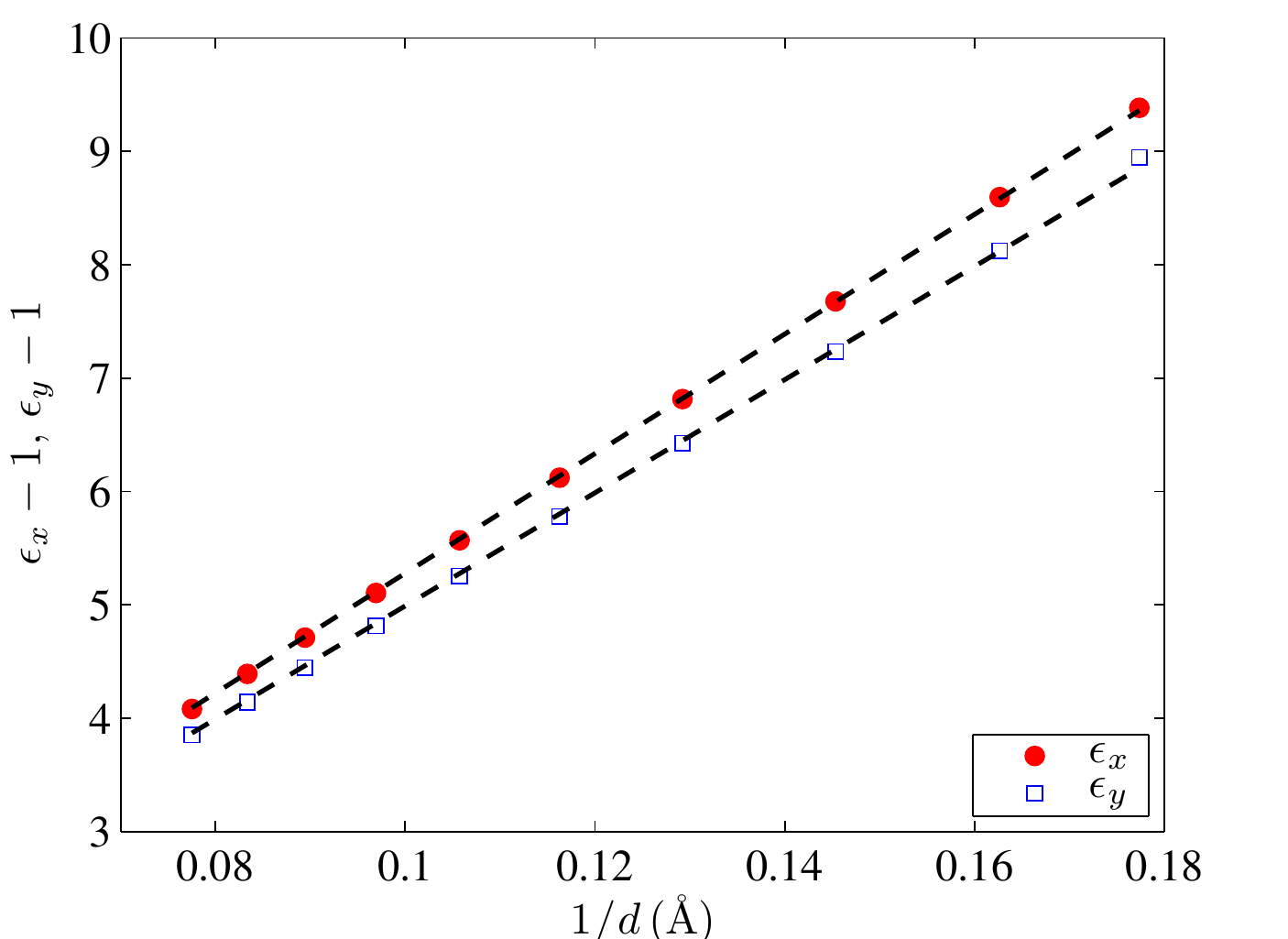}
\caption{Linear dependence of $\epsilon_{x,y}$ on the inverse interlayer distance $1/d$, see Eq.~\eqref{eqn:epsilon_xy}.}
\label{fig:Chi_Fit}
\end{figure}

The symmetry of the bulk black phosphorus unit cell was preserved as the inter-layer distance was increased up to three times the lattice parameter along
the $x$ direction.
The $x$ and $y$ components of the dielectric constant were obtained using the {\sc Quantum ESPRESSO} code.\cite{Giannozzi2009}
The exchange correlation energy was described by the generalized gradient approximation (GGA)
using the PBE functional.\cite{PBE}
The Kohn-Sham orbitals were expanded in a plane-wave basis with a cutoff energy of 70~Ry.
The Kohn-Sham states corresponding to the valence and conduction bands are shown in Fig.~\ref{fig:Map}. 
 For the dielectric tensor calculation, a rigid "scissors operator” shift of 0.72 eV was applied to the
 Kohn-Sham eigenvalues.
This corrects for the difference between the nearly vanishing PBE bandgap of bulk black phosphorus (80~meV)
and the value obtained by previous GW calculations.\cite{Tran2014tbg}
The Brillouin-zone (BZ) was sampled using a Monkhorst-Pack grid of 15$\times$40$\times$40 points 
along each of the primitive lattice vectors.\cite{monkhorst-prb-13-5188}
In this way, we obtain a linear dependence of $\epsilon_{x,y}$ on the inverse inter-layer distance,
with $\zeta_{xx}=4.20$~\AA\, and $\zeta_{yy}=3.97$~\AA, Fig.~\ref{fig:Chi_Fit}.  Since the values are fairly close, we use the average and set $\zeta = 4.1$ \AA. This yields $W \approx 48.6$ and $G\approx 13.6/\kappa^2$.

\begin{figure}
\includegraphics[width = 3in]{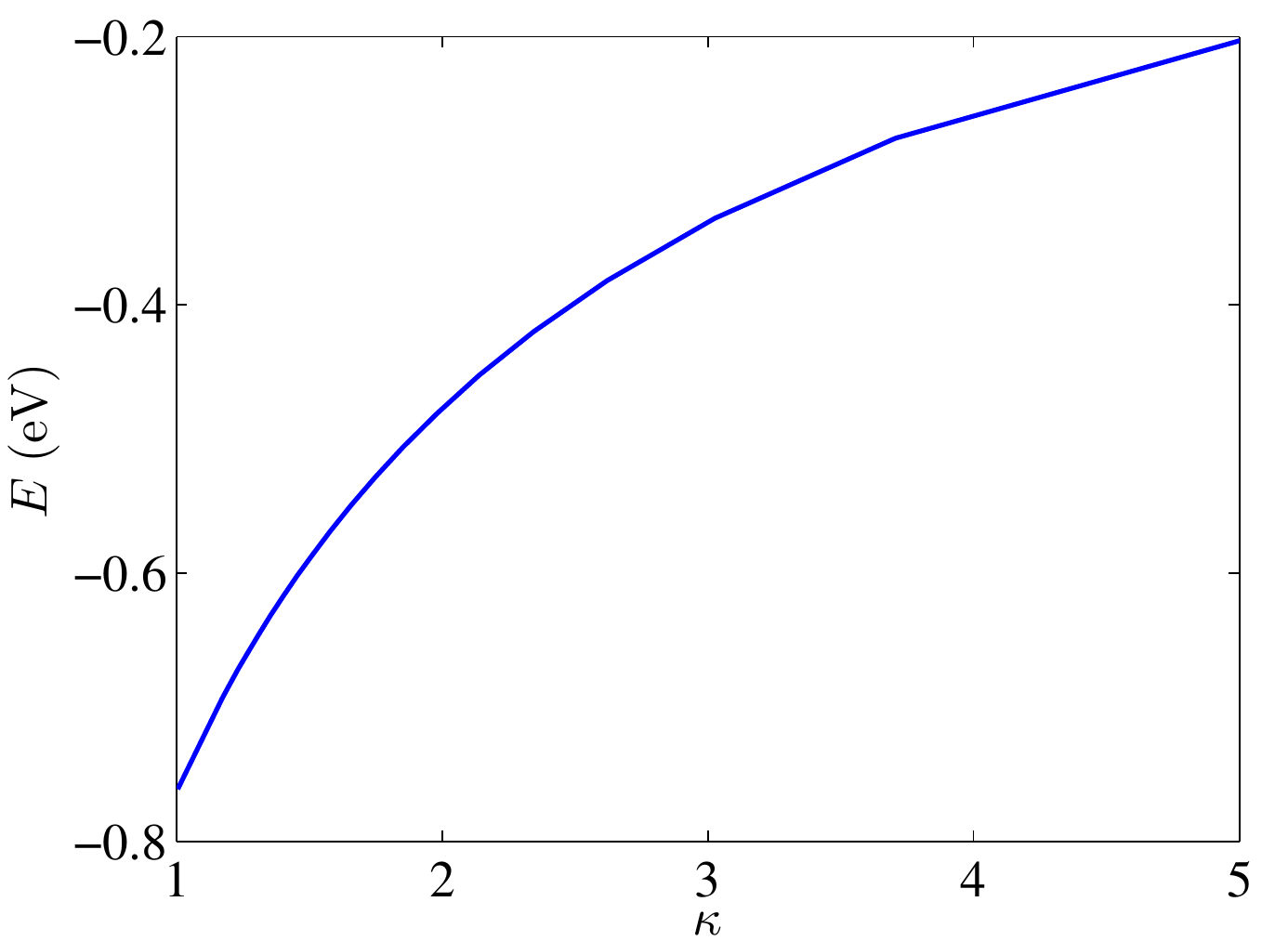}
\caption{Exciton binding energy as a function of $\kappa$.}
\label{fig:E_Kappa}
\end{figure}

It is now possible for us to determine the excitonic binding energy in phosphorene. Since the dependence of the interaction strength $G$ on the dielectric constant of the substrate is rather simple, we can obtain the binding energy as a function of $\kappa$. To do so, we compute the lowest excitonic energy for $\beta = 0.9$ for a range of $\kappa$'s between 1 and 5, as shown in Fig.~\ref{fig:E_Kappa}. For the case of isolated phosphorene, given by $\kappa = 1$, the binding energy is $0.76$ eV. This value is close to the one obtained from the first principles calculations in an earlier work~\cite{Tran2014tbg}. There, the authors reported the binding energy to be $0.8$ eV. With increasing $\kappa$, the lowest bound state becomes more shallow due to screening. In the case of phosphorene positioned on $\text{SiO}_2$, the exciton binding energy is close to $0.4$ eV, similar to the value reported in Ref.~\onlinecite{Castellanos_Gomez2014iac}.

\begin{figure}
\includegraphics[width = .75in]{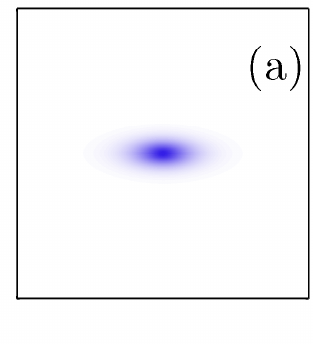}
\includegraphics[width = .75in]{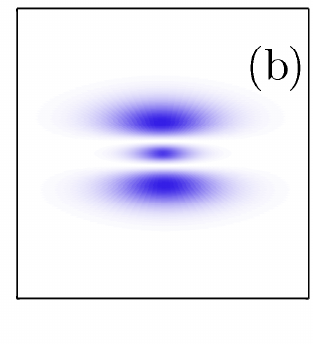}
\includegraphics[width = .75in]{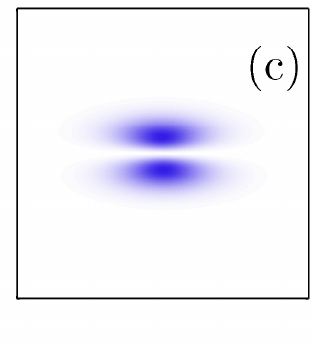}
\includegraphics[width = .75in]{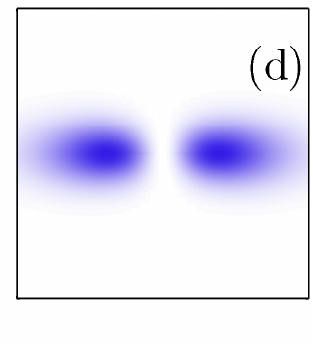}
\caption{Squared wavefunctions of phosphorene orbitals. From left to right: $1s$, $2s$, $2p_y$, and $2p_x$. The size of each frame is $100\times100$ \AA.}
\label{fig:BP_Orbitals}
\end{figure}

We can also compute the wavefunctions of phosphorene excitons, see Fig.~\ref{fig:BP_Orbitals}. At the first glance, it might appear strange that the wavefunctions are stretched in $x$ direction, in contradiction to the results shown in Figs.~\ref{fig:S_Orbitals}--\ref{fig:P_Orbitals}. However, one needs to keep in mind the change of variables in Eq.~\eqref{eqn:Scale_Var}. When we go back to the original real-space variables, the orbitals become stretched in the $x$ direction since the $x$ mass is much smaller than the $y$ mass. From Fig.~\ref{fig:BP_Orbitals}, we can see that the excitons are fairly large, spanning tens of Angstroms. This provides additional validation to our approach of using the continuum approximation in Eq.~\eqref{eqn:H_General}

\section{Conclusions}

Using a combination of the first principles calculations, Numerov method, and analytics, we study the general excitonic behavior of anisotropic 2D systems. We employ a modified electron-hole interaction which includes screening due to the 2D system itself, as well as due to the dielectric substrate. Our results show the dependence of the excitonic energies on both the interaction strength and the anisotropy parameter arising from the direction-dependent effective masses. Unlike the unscreened Coulomb case, the energy has a weaker, sub-quadratic dependence on the interaction strength with higher energy levels being more sensitive.

From our results, we compute the exciton binding energy for phosphorene. We see that our solution for the isolated monolayer agrees with the  earlier GW calculations~\cite{Tran2014tbg} and phosphorene on silicon dioxide is congruent with the value obtained using variational methods~\cite{Castellanos_Gomez2014iac}. The main advantage of our approach over the other two is the reduced requirement for the computational power compared to the GW and the applicability for higher energy levels where variational methods lose accuracy.

A.S.R. acknowledges DOE grant DE-FG02-08ER46512, ONR grant MURI N00014-09-1-1063. A.H.C.N. acknowledges NRF-CRP award ``Novel 2D materials with tailored properties: beyond graphene" (R-144-000-295-281). The DFT calculations were performed in the GRC computing facilities.
%

\end{document}